\documentclass[aps,prb,reprint,superscriptaddress,floatfix]{revtex4-2}

\usepackage{amsmath, amssymb}
\usepackage{bm}
\usepackage{graphicx}
\usepackage{xcolor}
\usepackage{dcolumn}
\usepackage{multirow}
\newcommand{\AF}{\text{AF}\,x^2\!-\!y^2}
\newcommand{\Fz}{\text{F}\,z^2}
\usepackage{xr-hyper}
\usepackage[hidelinks]{hyperref}
\hypersetup{colorlinks=true, urlcolor=blue, linkcolor=blue, citecolor=red}

\allowdisplaybreaks[1]

\begin{document}

\title{Spin-vibronic order in Ba$_2$MgReO$_6$}

\author{Zhishuo Huang}
\affiliation{Electrical Engineering and Computer Science, Technische Universit\"{a}t Berlin, Stra{\ss}e des 17. Juni 135, 10623 Berlin, Germany}

\author{Takumi Gengo}
\affiliation{Graduate School of Science and Engineering, Chiba University, 1-33 Yayoi-cho, Inage-ku, Chiba-shi, Chiba 263-8522, Japan}

\author{Teruki Matsuzaki}
\affiliation{Graduate School of Science and Engineering, Chiba University, 1-33 Yayoi-cho, Inage-ku, Chiba-shi, Chiba 263-8522, Japan}

\author{Liviu F. Chibotaru}
\affiliation{Theory of Nanomaterials Group, KU Leuven, Celestijnenlaan 200F, B-3001 Leuven, Belgium}

\author{Naoya Iwahara}
\email[]{naoya.iwahara@gmail.com}
\affiliation{Graduate School of Engineering, Chiba University, 1-33 Yayoi-cho, Inage-ku, Chiba-shi, Chiba 263-8522, Japan}

\begin{abstract}
Strong spin-orbit coupling in $5d$ ions produces multipolar orders that conventional probes barely detect.
In cubic $5d^1$ double perovskites, the spin-orbit quartet couples to Jahn-Teller vibrations, and resonant inelastic x-ray scattering (RIXS) has established the resulting spin-orbit-lattice entangled (vibronic) states [Phys. Rev. Lett. \textbf{133}, 036501 (2024)].
We proposed that these vibronic states order cooperatively and thereby explain the hidden quadrupole and canted magnetic phases of Ba$_2$MgReO$_6$ [Phys. Rev. B \textbf{107}, L220404 (2023)], but that proposal rested on a simplified model.
Here, we show that a spin-vibronic Hamiltonian derived from first principles, with a single scale factor for the elastic coupling, reproduces the ordered phases of Ba$_2$MgReO$_6$ with the same vibronic states that describe its RIXS spectra.
With this factor bringing the quadrupolar transition to 34.7 K, near the measured 33 K, the theory predicts the magnetic transition at 19.3 K, against the measured 18 K, and reproduces the antiferro distortion and the ferro elongation between the transitions.
These results support spin-vibronic order in Ba$_2$MgReO$_6$ and point to it as the origin of the hidden multipolar phases of $5d^1$ double perovskites.
\end{abstract}

\maketitle

\paragraph*{Introduction.}
In $4d$ and $5d$ transition-metal compounds, strong spin-orbit coupling acts together with electron correlation, entangles spin and orbital moments, and creates new quantum phases \cite{WitczakKrempa2014, Takayama2021, Takagi2019}.
Ions with less than half-filled $t_{2g}$ shells, such as $5d^1$ Re$^{6+}$ and Os$^{7+}$ and $5d^2$ Os$^{6+}$, keep orbitally degenerate spin-orbit ground states, and their compounds host multipolar orders that escape conventional magnetic probes \cite{Chen2024npj, Pourovskii2025}.

Cubic $5d^1$ double perovskites are the simplest multipolar case: their $5d^1$ octahedra form a face-centered cubic (fcc) lattice, and spin-orbit coupling turns the single $t_{2g}$ electron into a $\Gamma_8$ quartet with dipole, quadrupole, and octupole moments \cite{Chen2010}.
Osmates \cite{Erickson2007, Steele2011, Lu2017, Liu2018PRB, Liu2018, Willa2019}, rhenates \cite{Marjerrison2016, Hirai2019, Hirai2021, Ishikawa2021b, Gao2020, Woodward2022}, a tungstate \cite{Lee2021}, and Ta chlorides \cite{Ishikawa2019, Ishikawa2021, Tehrani2023} show multipolar orders, small ordered moments, or glassy states, and many of them undergo structural transitions, which points to an active role of the lattice.

Theories of multipolar order, in $d$- and $f$-electron systems alike, usually treat the electrons alone, although the lattice also takes part \cite{Santini2009, KaplanVekhter1995}.
A perturbative treatment of the local electron-phonon coupling yields a lattice-mediated quadrupole interaction \cite{Sugihara1959}.
Beyond this weak-coupling regime, the Jahn-Teller effect shapes the local states, and spin-orbit coupling only partially suppresses it in the $t_{2g}^1$ configuration \cite{Thorson1968, Xu2016, Iwahara2018, Streltsov2020}.

In the dynamic Jahn-Teller regime, the lattice remains fully quantum: we showed that in cubic double perovskites the coupling of the quartet to the $E_g$ and $T_{2g}$ vibrations, in the intermediate-coupling regime, turns the local states into spin-orbit-lattice entangled (vibronic) states \cite{Iwahara2018, Iwahara2024JPSJ}.
Resonant inelastic x-ray scattering (RIXS) probes the dynamic Jahn-Teller effect directly.
It shows Jahn-Teller activity in Ba$_2$MgReO$_6$ \cite{Zivkovic2024} and Ba$_2$NaOsO$_6$ \cite{Agrestini2024}, and the theory based on vibronic states reproduces the RIXS spectra of Re double perovskites \cite{Frontini2024, Iwahara2025PRB, Matsuzaki2026}.
Because the vibronic stabilization far exceeds the intersite interactions in the family of $5d^1$ double perovskites, the vibronic states survive in the crystal as local building blocks that add vibrational degrees of freedom to the multipolar interactions and should order cooperatively at low temperatures \cite{Iwahara2023}.

\begin{figure}[tb]
\includegraphics[width=\linewidth]{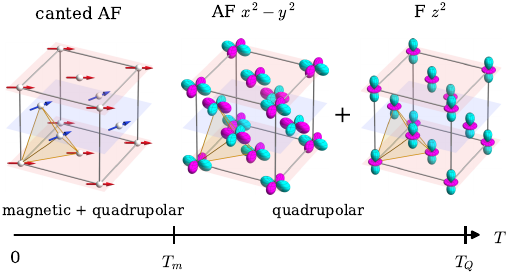}
\caption{\label{Fig:order}
Ordered phases of Ba$_2$MgReO$_6$ in the conventional cubic cell; the shaded planes mark the (001) layers normal to the modulation axis $z$, and the orange tetrahedron connects four nearest-neighbor Re sites.
Between $T_m$ and $T_Q$ the antiferro $x^2-y^2$ and ferro $z^2$ quadrupoles coexist (cyan and magenta: positive and negative parts of the quadrupole shapes).
Below $T_m$ the moments order in a canted antiferromagnet: in alternate layers (red and blue arrows on the planes of the same color) they deviate by $\pm\phi_m$ from the net moment along $[110]$, while the quadrupole order persists.
}
\end{figure}

Ba$_2$MgReO$_6$ is one of the best systems to establish the ordering of vibronic states, because it clearly shows both lattice and magnetic order (Fig. \ref{Fig:order}): an antiferro $x^2-y^2$ ($\AF$) distortion with a ferro $z^2$ ($\Fz$) elongation appears at $T_Q = 33$ K, and a canted antiferromagnet at $T_m = 18$ K \cite{Hirai2019, Hirai2020, Soh2024}.
We showed that the dynamic Jahn-Teller effect on the sites gives rise to a quadrupolar phase between $T_m$ and $T_Q$ in which the $\AF$ and $\Fz$ orders coexist, and that this spin-vibronic order explains both observed phases \cite{Iwahara2023}.
Electronic multipolar models obtain this coexistence through the admixture of the excited $\Gamma_7$ doublet, but with intersite interactions that reproduce the observed transition temperatures the induced ferro $z^2$ remains far smaller than observed \cite{Svoboda2021, Kubo2023}.
Classical Monte Carlo simulations show that intersite correlations can change the ordered phases \cite{Cong2026, Hart2026}.

Whether first-principles interactions between the vibronic states seen by RIXS reproduce the ordered phases of Ba$_2$MgReO$_6$ at the observed temperatures has remained open.
First-principles calculations have addressed these phases mainly within the static Jahn-Teller regime, where they describe the distorted low-temperature phases of Ba$_2$NaOsO$_6$ and Ba$_2$MgReO$_6$ \cite{Cong2020, FioreMosca2021, Merkel2024}.
For Ba$_2$MgReO$_6$, density functional theory (DFT) calculations at zero temperature obtain the $\AF$ and $\Fz$ quadrupoles in the magnetically ordered state \cite{MansouriTehrani2021} and show that the Jahn-Teller coupling stabilizes the $\AF$ order \cite{Martinelli2026}.
A first-principles model with quantum Jahn-Teller phonons, its vibronic basis truncated at one phonon, yields $\Fz$ only below $T_m$ \cite{FioreMosca2024}.
Thus no first-principles study has obtained the $\Fz$ elongation between $T_m$ and $T_Q$ with quantum-mechanical lattice vibrations.

Here, we show that the spin-vibronic order \cite{Iwahara2023} emerges from a Hamiltonian built from first principles that keeps all low-energy vibronic states of each site.
Its single-site part is the vibronic Hamiltonian that reproduces the RIXS spectra \cite{Matsuzaki2026}, so that one set of vibronic states describes both the RIXS spectra and the ordered phases.
We solve the model in the mean-field approximation and correct the transition temperatures with the Onsager reaction field.
With the elastic coupling halved to bring $T_Q$ near experiment, the theory predicts $T_m = 19.3$ K against the measured 18 K and reproduces the observed quadrupolar and magnetic orders.

\paragraph*{Spin-vibronic Hamiltonian.}
We describe Ba$_2$MgReO$_6$ by the spin-orbit-coupled Re $t_{2g}$ electrons and the Jahn-Teller-active $E_g$ and $T_{2g}$ vibrations of the ReO$_6$ octahedra, both treated quantum mechanically:
\begin{align}
 \hat{H} &= \sum_i \hat{H}^i_\text{DJT} + \sum_{\langle ij \rangle} \left( \hat{H}^{ij}_\text{ex} + \hat{H}^{ij}_\text{el} \right).
 \label{Eq:H}
\end{align}
Here, $\hat{H}^i_\text{DJT}$ is the dynamic Jahn-Teller Hamiltonian of site $i$, $\hat{H}^{ij}_\text{ex}$ the exchange interaction between the $t_{2g}$ electrons of sites $i$ and $j$, and $\hat{H}^{ij}_\text{el}$ the elastic coupling between their octahedral distortions; the sum over pairs $\langle ij \rangle$ runs over the nearest to fourth-nearest neighbors of the fcc Re lattice, for both the exchange and the elastic coupling.
Equation (\ref{Eq:H}) retains the major part of the cooperative Jahn-Teller Hamiltonian [Ref. \cite{Kanamori1960}, Eq. (11)]: the Jahn-Teller-active vibrations of each octahedron are quantum dynamical variables, and the other lattice modes enter through the elastic coupling; the coupling to the uniform strain is omitted (see below).

The single-site Hamiltonian contains the spin-orbit coupling, the harmonic vibrations, and the linear and quadratic vibronic couplings \cite{Iwahara2018, Matsuzaki2026}:
\begin{align}
 \hat{H}^i_\text{DJT} &= \lambda \tilde{\bm{l}} \cdot \hat{\bm{s}}
 + \sum_{k} \frac{\hslash\omega_k}{2} \left( \hat{p}_k^2 + \hat{q}_k^2 \right)
 + \sum_{k} g_k \hslash\omega_k \hat{q}_k \hat{C}_k
 \nonumber\\
 &\quad + \sum_{\gamma = \theta, \epsilon} \left( \hslash\omega_E  \frac{w_{E}}{2} [\hat{q}_E^2]_\gamma + \hslash \omega_{T_2} \frac{w_{T_2}}{2} [\hat{q}_{T_2}^2]_\gamma \right) \hat{C}_\gamma.
 \label{Eq:Hloc}
\end{align}
The index $k$ runs over the components $\theta, \epsilon$ of the $E_g$ mode and $\xi, \eta, \zeta$ of the $T_{2g}$ mode, with the linear coupling constants $g_k = g_E$ and $g_{T_2}$, respectively; $w_E$ and $w_{T_2}$ are the quadratic coupling constants.
$\hat{q}_k$ and $\hat{p}_k$ are the dimensionless normal coordinates and momenta, $\hat{C}_k$ the $t_{2g}$ orbital operators of the same symmetry \cite{Iwahara2018}, and $[\hat{q}_\Gamma^2]_\gamma$ ($\Gamma = E, T_2$) the $E_g$ components of the symmetrized squares of the coordinates.
The axes $x$, $y$, $z$ are the cubic axes $[100]$, $[010]$, $[001]$; positive $q_\theta$ elongates the octahedron along $z$, and $\tilde{\bm{l}}$ is the atomic $d$ orbital angular momentum projected onto the $t_{2g}$ shell [Ref. \cite{Sugano1970}, Sec. 7.1.1].

The low-energy vibronic states of each site span the space on which all intersite interactions act.
As in our previous works \cite{Iwahara2018, Frontini2024, Iwahara2025PRB, Matsuzaki2026}, we obtain them by numerically diagonalizing Eq. (\ref{Eq:Hloc}) in the basis of products of the six $t_{2g}$ spin-orbitals and the harmonic-oscillator states of the $E_g$ and $T_{2g}$ modes, with up to 12 vibrational quanta.
We keep the 20 lowest vibronic states $|\Psi^i_n\rangle$ of each site $i$, with energies $E_n$: the ground vibronic quartet, the first excited doublet at 32.7 meV, and the 14 states between 47.8 and 54.8 meV\footnote{The next vibronic level lies at 75.6 meV, so the truncation falls in a gap and splits no multiplet. Calculations with all 48 states up to 99 meV at several temperatures change the transition temperatures by less than 1\%, the order parameters at $T \to 0$ by at most 3\%, and $\Fz$ between $T_m$ and $T_Q$ by 7\%.}.
With the projector $\hat{\mathcal{P}}^i = \sum_{n=1}^{20} |\Psi^i_n\rangle \langle \Psi^i_n|$, the single-site Hamiltonian of Eq. (\ref{Eq:H}) becomes
\begin{align}
 \hat{\mathcal{H}}^i_\text{DJT} &= \hat{\mathcal{P}}^i \hat{H}^i_\text{DJT} \hat{\mathcal{P}}^i = \sum_n E_n |\Psi^i_n\rangle \langle \Psi^i_n| .
 \nonumber
\end{align}

\paragraph*{Superexchange interaction.}
The superexchange interaction is the second-order effective Hamiltonian of the electron transfer $\hat{H}^{ij}_t = \sum_{ab} t^{ij}_{ab} \hat{d}^{i\dagger}_a \hat{d}^j_b + \text{H.c.}$ between two Re sites \cite{Anderson1959}, where $\hat{d}^{i\dagger}_a$ ($\hat{d}^i_a$) creates (annihilates) an electron in the $t_{2g}$ spin-orbital $a$ of site $i$, and $t^{ij}_{ab}$ is diagonal in spin and given by the orbital transfer integrals (Table \ref{Table:param}):
\begin{align}
 \hat{H}^{ij}_\text{ex} &= - Z_B^2 \hat{H}^{ij}_t \left( \sum_m \frac{|m\rangle\langle m|}{\Delta_m} \right) \hat{H}^{ij}_t ,
 \label{Eq:Hexfull}
\end{align}
where $Z_B^2$ accounts for the frequency dependence of the screening (see below) and $|m\rangle$ are the $d^0 d^2$ intermediate multiplets at excitation energies $\Delta_m$.
We obtain $|m\rangle$ and $\Delta_m$ by diagonalizing the $d^2$ ion exactly with the Kanamori interaction ($U$, $U'$, $J_\text{H}$) \cite{Kanamori1963} and the spin-orbit coupling, and include the intersite repulsion $V$.
Projected onto the low-energy vibronic states of the two sites by $\hat{\mathcal{P}}^{ij} = \hat{\mathcal{P}}^i \hat{\mathcal{P}}^j$, it becomes
\begin{align}
 \hat{\mathcal{H}}^{ij}_\text{ex} &= \hat{\mathcal{P}}^{ij} \hat{H}^{ij}_\text{ex} \hat{\mathcal{P}}^{ij} = \sum_{aa'bb'} M^{ij}_{aa',bb'} \hat{\mathcal{X}}^i_{aa'} \hat{\mathcal{X}}^j_{bb'} ,
 \label{Eq:Hex}
\end{align}
with the projected electronic operators $\hat{\mathcal{X}}^i_{ab} = \hat{\mathcal{P}}^i \left( |a\rangle\langle b| \otimes \hat{1}_\text{vib} \right) \hat{\mathcal{P}}^i$ of site $i$, where $|a\rangle$ and $|b\rangle$ run over the six $t_{2g}$ spin-orbitals.
The projected operators $\hat{\mathcal{X}}^i_{ab}$ contain the reduction of every electronic multipole by the vibronic coupling \cite{Ham1965}.

\paragraph*{Elastic coupling.}
The distortions of different octahedra interact through the lattice,
\begin{align}
 \hat{\mathcal{H}}^{ij}_\text{el} &= f_\text{el} \sum_{kl} D^{ij}_{kl} \hat{\mathcal{P}}^{ij} \hat{q}^i_k \hat{q}^j_l \hat{\mathcal{P}}^{ij}
 = f_\text{el} \sum_{kl} D^{ij}_{kl} \hat{\mathcal{Q}}^i_k \hat{\mathcal{Q}}^j_l ,
 \label{Eq:Hel_1}
\end{align}
with the elastic coupling constants $D^{ij}_{kl}$, a scale factor $f_\text{el}$, and the projected coordinates $\hat{\mathcal{Q}}^i_k = \hat{\mathcal{P}}^i \hat{q}^i_k \hat{\mathcal{P}}^i$.
We construct $D^{ij}_{kl}$ from the static lattice Green's function [Ref. \cite{Chibotaru1994}, Sec. 3] of the dynamical matrix, which contains no uniform strain.

\begin{table}[tb]
\caption{\label{Table:param}
Parameters of the Hamiltonian (\ref{Eq:H}) of Ba$_2$MgReO$_6$ (energies in meV).
Single site: parameters of Eq. (\ref{Eq:Hloc}) from Ref. \cite{Matsuzaki2026}, $\kappa$ from Ref. \cite{Iwahara2018}, and on-site Coulomb parameters from the constrained random-phase approximation (cRPA).
Two sites, nearest-neighbor $[\frac12 \frac12 0]$ and next-nearest-neighbor $[001]$ bonds: transfer integrals from PBE Wannier functions, intersite repulsion from cRPA, and elastic couplings from the lattice Green's function \cite{Chibotaru1994} before scaling by $f_\text{el}$; the parameters of all bonds of the four neighbor shells and the computational settings are given in the Supplemental Material \cite{SM}.
}
\begin{ruledtabular}
\begin{tabular}{llll}
\multicolumn{4}{l}{Single site} \\
$U$, $U'$ & 2782, 2183 & $J_\text{H}$ & 252 \\
$\lambda$ & 321 & $\kappa$ & 0.56 \\
$\hslash\omega_E$, $\hslash\omega_{T_2}$ & 63.1, 55.6 & $g_E$, $g_{T_2}$ & 1.332, 0.71 \\
$w_{E}$, $w_{T_2}$ & 0.23, $-0.21$ & & \\
\hline
\multicolumn{4}{l}{Two sites, $[\frac12 \frac12 0]$} \\
$t_{xy,xy}$ & $-113.7$ & $t_{yz,yz} = t_{zx,zx}$ & 25.4 \\
$t_{yz,zx}$ & 14.4 \\
$V$ & 526 \\
$D_{\theta\theta}$, $D_{\epsilon\epsilon}$ & 0.71, $-2.42$ & $D_{\theta\zeta}$, $D_{\zeta\zeta}$ & $-1.47$, 1.49 \\
$D_{\xi\xi} = D_{\eta\eta}$ & $-0.57$ & $D_{\xi\eta}$ & $-0.22$ \\
\hline
\multicolumn{4}{l}{Two sites, $[001]$} \\
$t_{yz,yz} = t_{zx,zx}$ & 3.6 & $t_{xy,xy}$ & $-0.5$ \\
$D_{\theta\theta}$, $D_{\epsilon\epsilon}$ & $-0.08$, $-0.45$ & $D_{\xi\xi} = D_{\eta\eta}$, $D_{\zeta\zeta}$ & $-0.94$, $-1.09$ \\
\end{tabular}
\end{ruledtabular}
\end{table}

\paragraph*{First-principles calculations.}
Equation-of-motion coupled-cluster calculations on embedded ReO$_6$ clusters give the single-site parameters of Eq. (\ref{Eq:Hloc}) (Table \ref{Table:param}), which agree within 5\% with those extracted from the RIXS spectra \cite{Matsuzaki2026}.
We compute the remaining parameters for the experimental cubic structure \cite{Marjerrison2016} with the Perdew-Burke-Ernzerhof (PBE) functional \cite{HZSPerdew1996generalized}.
For the transfer integrals and the static Coulomb parameters, we compute the band structure of the conventional cell with Quantum ESPRESSO \cite{HZSGiannozzi2017advanced} and optimized norm-conserving Vanderbilt pseudopotentials \cite{HZSHamann2013oprimized}, construct maximally localized Wannier functions \cite{Marzari2012} of the 12 Re $t_{2g}$ bands, and evaluate the constrained random-phase approximation (cRPA) \cite{Aryasetiawan2004} at zero frequency with RESPACK \cite{Nakamura2021RESPACK}; we omit the direct intersite quadrupole Coulomb interaction, which cRPA on metallic DFT orbitals does not give reliably.
For the frequency-dependent interaction $U(\omega)$, we use VASP \cite{Kresse1996} with projector augmented-wave potentials \cite{Blochl1994, Kresse1999} and Wannier90 \cite{Pizzi2020}, and compute the cRPA interaction at real frequencies between 0.5 and 50 eV.
The elastic couplings $D^{ij}_{kl}$ follow from the frozen-phonon force constants of Ref. \cite{Huang2024} through the lattice Green's function.

The factor $Z_B^2$ in Eq. (\ref{Eq:Hexfull}) accounts for the frequency dependence of the screened Coulomb interaction.
The static cRPA interaction treats the screening as instantaneous, whereas the screening builds up only at the interband and plasmon frequencies, 2.5--30 eV.
The electrons then hop with their screening clouds, which reduces every transfer integral by a factor $Z_B$ determined by the imaginary part of $U(\omega)$ \cite{Casula2012PRL}.
Our cRPA $U(\omega)$ gives $Z_B = 0.81$, so that the superexchange is reduced by the factor $Z_B^2 = 0.66$.

The PBE force constants are those of a metal, because DFT cannot represent the insulating spin-vibronic state of Ba$_2$MgReO$_6$.
We therefore scale the elastic couplings $D^{ij}_{kl}$ by a factor $f_\text{el}$, the only adjusted parameter, and fix $f_\text{el} = 1/2$ by $T_Q$.
A related example is UO$_2$, a correlated insulator in which the on-site Coulomb repulsion strongly raises the zone-boundary frequencies of the optical branches \cite{Zhou2022} and thereby reduces the elastic coupling.

\paragraph*{Magnetic and multipolar moments.}
The magnetic moment of site $i$ is
\begin{align}
 \hat{\bm{\mathcal{M}}}^i &= -\mu_\text{B} \hat{\mathcal{P}}^i \left( \kappa \tilde{\bm{l}}^i + g_e \hat{\bm{s}}^i \right) \hat{\mathcal{P}}^i ,
 \label{Eq:m}
\end{align}
where $g_e$ is the electron $g$ factor, and $\kappa$ the covalent reduction factor of the orbital angular momentum \cite{Ahn2017}.
We take $\kappa = 0.56$ of the isoelectronic Os double perovskites \cite{Iwahara2018}; $\kappa$ enters only the moment and not the ordering.
With this $\kappa$, the susceptibility of the single-site vibronic states, fitted to the Curie-Weiss law between 80 and 350 K as in the experiment, gives the effective moment 0.71 $\mu_\text{B}$, close to the measured 0.68 $\mu_\text{B}$ \cite{Hirai2019}.

We describe the electronic order by the multipoles of the pseudo-angular momentum $\hat{\bm{J}}^i = \hat{\bm{s}}^i - \tilde{\bm{l}}^i$; $\hat{\bm{J}}^i$ is a proper angular momentum because $-\tilde{\bm{l}}^i$ equals the $l = 1$ momentum of the $p$ orbitals $(x, y, z)$ in the basis $(yz, zx, xy)$ [Ref. \cite{Sugano1970}, Sec. 7.1.1], and it labels the ground quartet as $j = 3/2$. 
With $z$ as the quantization axis, the electronic multipoles are, up to constant factors, those of Ref. \cite{Shiina1998},
\begin{align}
 \hat{O}_{20} &= \tfrac{1}{\sqrt{3}} \left( 3\hat{J}_z^2 - \hat{\bm{J}}^2 \right), &
 \hat{O}_{22} &= \hat{J}_x^2 - \hat{J}_y^2, \nonumber\\*
 \hat{O}_{\alpha\beta} &= \{\hat{J}_\alpha, \hat{J}_\beta\}, &
 \hat{T}_{xyz} &= \overline{\hat{J}_x \hat{J}_y \hat{J}_z}, \nonumber\\*
 \hat{T}^{\Gamma_4}_x &= \hat{J}_x^3 - \tfrac{1}{10} \{3\hat{\bm{J}}^2 - 1, \hat{J}_x\}, &
 \hat{T}^{\Gamma_5}_x &= \tfrac12 \{\hat{J}_x, \hat{J}_y^2 - \hat{J}_z^2\},
 \label{Eq:multipoles}
\end{align}
with $\alpha\beta = yz, zx, xy$, and cyclically for $y$ and $z$ in $\hat{T}^{\Gamma_4}_x$ and $\hat{T}^{\Gamma_5}_x$, where the bar denotes symmetrization, $\{\,,\}$ the anticommutator, and the superscripts $\Gamma_4$ and $\Gamma_5$ the irreducible representations of the octupoles in $O_h$.
Built from $\hat{\bm{J}}^i$ and projected as $\hat{\bm{\mathcal{M}}}^i$ in Eq. (\ref{Eq:m}), they give the moments of the vibronic states of site $i$, $\hat{\mathcal{J}}^i_\alpha = \hat{\mathcal{P}}^i (\hat{J}^i_\alpha \otimes \hat{1}_\text{vib}) \hat{\mathcal{P}}^i$, $\hat{\mathcal{O}}^i_{20} = \hat{\mathcal{P}}^i (\hat{O}^i_{20} \otimes \hat{1}_\text{vib}) \hat{\mathcal{P}}^i$, and so on.

\paragraph*{Mean-field and reaction-field theory.}
We first solve Eq. (\ref{Eq:H}) in the single-site mean-field approximation on the four-site conventional cubic cell, which accommodates both the quadrupole and the canted magnetic orders \cite{Iwahara2023}.
We collect the five $\hat{\mathcal{Q}}^i_k$ and 36 Hermitian combinations of the $\hat{\mathcal{X}}^i_{ab}$ of site $i$ into $\hat{\mathcal{A}}^i_\mu$ and write every bond of Eqs. (\ref{Eq:Hex}) and (\ref{Eq:Hel_1}) as $\sum_{\mu\nu} W^{ij}_{\mu\nu} \hat{\mathcal{A}}^i_\mu \hat{\mathcal{A}}^j_\nu$.
Minimizing the free energy of a product state gives the mean-field Hamiltonian of site $i$, which acts in its 20-dimensional vibronic space,
\begin{align}
 \hat{\mathcal{H}}^i_\text{MF} &= \hat{\mathcal{H}}^i_\text{DJT} + \sum_{j(\ne i)} \sum_{\mu\nu} W^{ij}_{\mu\nu} \hat{\mathcal{A}}^i_\mu \langle \hat{\mathcal{A}}^j_\nu \rangle .
 \label{Eq:MF}
\end{align}
Its eigenstates $|\Phi^i_\tau\rangle$, with energies $\varepsilon^i_\tau$ and thermal populations $p^i_\tau \propto e^{-\varepsilon^i_\tau/k_\text{B}T}$, give the expectation values $\langle \hat{\mathcal{A}}^i_\mu \rangle = \sum_\tau p^i_\tau \langle \Phi^i_\tau | \hat{\mathcal{A}}^i_\mu | \Phi^i_\tau \rangle$.
We iterate Eq. (\ref{Eq:MF}) from random initial states to self-consistency and keep the solution of lowest free energy.

The single-site mean-field approximation neglects the correlated fluctuations of neighboring sites and therefore overestimates $T_Q$ and $T_m$ [Ref. \cite{Yosida1996}, Sec. 7.1].
We correct the transition temperatures with the Onsager reaction-field (ORF) theory \cite{BroutThomas1967}, the random-phase approximation (RPA) for all coupled operators, supplemented by a reaction field.
At each temperature $T$, the RPA gives the static linear response of the operators $\hat{\mathcal{A}}^s_\mu$ of the four sublattices $s$ of the conventional cell (Fig. \ref{Fig:order}) to a field of wave vector $\bm{q}$.
We obtain it by linearizing the mean-field equations about a reference state at the same $T$ [Ref. \cite{JensenMackintosh1991}, Sec. 3.5],
\begin{align}
 \chi^{\bm{q}} &= \left[ 1 + \chi_0 \left( W^{\bm{q}} - \Lambda \right) \right]^{-1} \chi_0 ,
 \label{Eq:chiORF}
\end{align}
where $W^{\bm{q}}$ is the Fourier transform of $W^{ij}$, 
$\chi_0$ is the static single-site susceptibility, 
and $\Lambda$ is the reaction field. 
With the mean-field solutions of sublattice $s$, the block $\chi^s_0$ of $\chi_0$ is [Ref. \cite{JensenMackintosh1991}, Sec. 3.3]
\begin{align}
 \chi^s_{0,\mu\nu} &= \sum_{\tau\tau'} -\frac{p^s_\tau - p^s_{\tau'}}{\varepsilon^s_\tau - \varepsilon^s_{\tau'}} \langle \Phi^s_\tau | \delta\hat{\mathcal{A}}^s_\mu | \Phi^s_{\tau'} \rangle \langle \Phi^s_{\tau'} | \delta\hat{\mathcal{A}}^s_\nu | \Phi^s_\tau \rangle ,
 \label{Eq:chi0}
\end{align}
where $\delta\hat{\mathcal{A}}^s_\mu = \hat{\mathcal{A}}^s_\mu - \langle \hat{\mathcal{A}}^s_\mu \rangle$.
Degenerate pairs, including $\tau = \tau'$, enter as the limit $\varepsilon^s_{\tau'} \to \varepsilon^s_\tau$ and give the Curie response, and the other pairs give the van Vleck response.

The reaction field shifts the divergence of Eq. (\ref{Eq:chiORF}) from the mean-field transition to a lower temperature by removing the self-interaction of each site \cite{Onsager1936, BroutThomas1967}.
In the RPA alone, $\Lambda = 0$, a site responds to the field that its own fluctuations induce on its neighbors, and Eq. (\ref{Eq:chiORF}) diverges at the mean-field transition.
The reaction field $-\Lambda^s \delta\langle \hat{\mathcal{A}}^s \rangle$ on each sublattice $s$ cancels this response.

We fix $\Lambda^s$ by a local sum rule, which we derive here for operators that commute with the Hamiltonian, taking the multipole moments as fixed [Ref. \cite{BroutThomas1967}, Eqs. (23)--(25)].
For such operators the static response to a field $\propto e^{i\bm{q}\cdot\bm{R}^i}$ on the operators $\hat{\mathcal{A}}^i_\nu$ of sublattice $s$ equals the thermal fluctuation,
\begin{align}
 \chi^{ss,\bm{q}}_{\mu\nu} &= \frac{1}{k_\text{B}T} \langle \delta\hat{\mathcal{A}}^{s,\bm{q}}_\mu \delta\hat{\mathcal{A}}^{s,-\bm{q}}_\nu \rangle_{\hat{H}} ,
 \label{Eq:FT}
\end{align}
where $\langle \cdots \rangle_{\hat{H}}$ is the thermal average with $\hat{H}$ of Eq. (\ref{Eq:H}), with respect to which $\delta$ is taken here, $\hat{\mathcal{A}}^{s,\bm{q}}_\mu = N_q^{-1/2} \sum_{i \in s} e^{-i\bm{q}\cdot\bm{R}^i} \hat{\mathcal{A}}^i_\mu$ and $N_q$ is the number of sites of sublattice $s$.
Averaging Eq. (\ref{Eq:FT}) over $\bm{q}$ leaves the on-site fluctuation $\langle \delta\hat{\mathcal{A}}^i_\mu \delta\hat{\mathcal{A}}^i_\nu \rangle_{\hat{H}}/k_\text{B}T$ of a site $i$ of $s$.
For fixed moments, the on-site fluctuation does not depend on the correlations with the neighbors and keeps its single-site value $\langle \delta\hat{\mathcal{A}}^s_\mu \delta\hat{\mathcal{A}}^s_\nu \rangle$ in the reference state.
Because the operators commute with $\hat{\mathcal{H}}^s_\text{MF}$, only the Curie terms of Eq. (\ref{Eq:chi0}) remain, $\chi^s_{0,\mu\nu} = \langle \delta\hat{\mathcal{A}}^s_\mu \delta\hat{\mathcal{A}}^s_\nu \rangle/k_\text{B}T$, and the two results give the local sum rule
\begin{align}
 \frac{1}{N_q} \sum_{\bm{q}} \chi^{ss,\bm{q}} &= \chi^s_0.
 \label{Eq:sumrule}
\end{align}
The RPA violates Eq. (\ref{Eq:sumrule}), and $\Lambda^s$ restores it at each temperature.
We impose Eq. (\ref{Eq:sumrule}) in the same form on our noncommuting operators, whose static susceptibilities include van Vleck terms, as an approximation.

We solve Eq. (\ref{Eq:sumrule}) for the symmetric matrices $\Lambda^s$ by Newton's method at each $T$ and locate the transition where $1 + \chi_0 (W^{\bm{q}} - \Lambda)$ first becomes singular.
The reference state is the cubic paramagnet for $T_Q$ and the quadrupole-ordered mean-field state with the magnetic (time-odd) fields set to zero for $T_m$, for which Eq. (\ref{Eq:sumrule}) is an approximation beyond its derivation for a disordered state \cite{BroutThomas1967}.
The ORF theory fixes only $T_Q$ and $T_m$.
We take the order parameters from the mean field and plot them on a temperature axis that maps 0, $T_m$, and $T_Q$ of the mean field [Table \ref{Table:results}, in parentheses] piecewise linearly onto those of the ORF theory.

\begin{table}[tb]
\caption{\label{Table:results}
Calculated and measured transition temperatures and order parameters.
Theory with $Z_B^2 = 0.66$ and $f_\text{el} = 1/2$: transition temperatures from the ORF theory (mean field in parentheses), order parameters from the mean field at $T \to 0$, their fractions just above $T_m$ on the ORF scale, and $\mu_\text{eff}$ from the single-site susceptibility (80--350 K).
}
\begin{ruledtabular}
\begin{tabular}{lcc}
 & Theory & Exp. \\
\hline
$T_Q$ (K) & 34.7 (57.5) & 33 \cite{Hirai2019} \\
$T_m$ (K) & 19.3 (29.5) & 18 \cite{Hirai2019} \\
$\AF$ (\AA) & 0.043 & 0.044(5)\footnotemark[1] \cite{Hirai2020} \\
$\Fz$ (\AA) & 0.018 & 0.01(1)\footnotemark[1] \cite{Hirai2020} \\
$m_\text{net}$ ($\mu_\text{B}$) & 0.33 & 0.46\footnotemark[2] \cite{Hirai2019} \\
$\mu_\text{eff}$ ($\mu_\text{B}$) & 0.71 & 0.68 \cite{Hirai2019} \\
$\Fz(T_m^+)/\Fz(0)$ & 0.41 & 0.70\footnotemark[3] \cite{Soh2024} \\
AF $\hat{\mathcal{O}}_{22}(T_m^+)/\hat{\mathcal{O}}_{22}(0)$ & 0.98 & 0.69\footnotemark[3] \cite{Soh2024} \\
\end{tabular}
\footnotetext[1]{At 6 K, on a crystal with three tetragonal domains \cite{Muroi2025}.}
\footnotetext[2]{Inferred from the magnetization at 7 T and 2 K, assuming one magnetic domain per tetragonal domain; theory at zero field.}
\footnotetext[3]{Theory at 19.6 K, experiment at 20 and 19 K: the $(10,0,0)$ splitting and the square root of the $(5,5,0)$ intensity at the quadrupolar resonance (Fig. \ref{Fig:tdep}).}
\end{ruledtabular}
\end{table}

\begin{figure*}[tb]
\includegraphics[width=\textwidth]{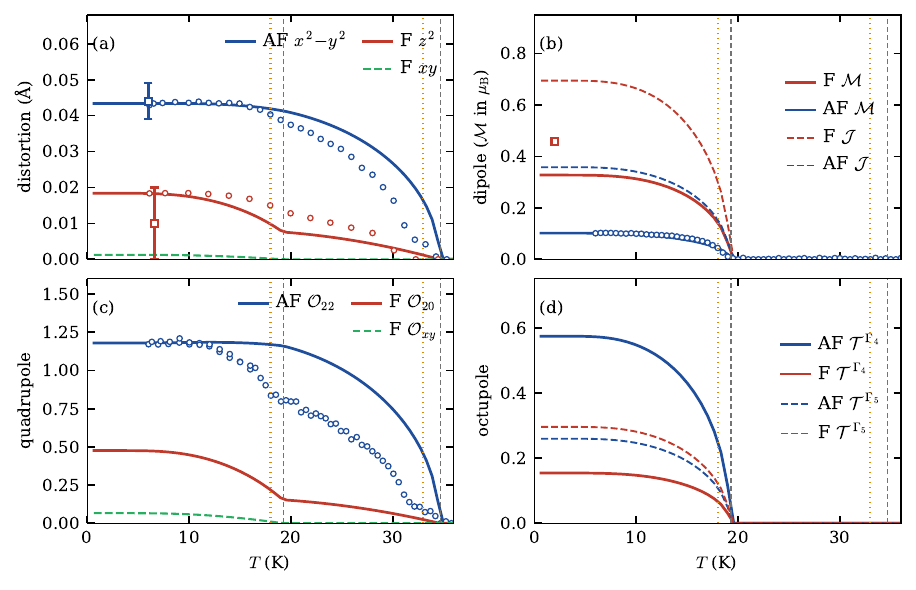}
\caption{
Order parameters on cooling with $Z_B^2 = 0.66$ and $f_\text{el} = 1/2$ (Table \ref{Table:results}); F and AF denote the ferro (uniform) and antiferro (staggered) components.
Mean-field values on the ORF temperature scale (piecewise linear map); dashed lines mark $T_Q = 34.7$ K and $T_m = 19.3$ K, dotted lines the measured 33 and 18 K \cite{Hirai2019}.
Circles: x-ray scans of Ref. \cite{Soh2024}, the square roots of the background-subtracted intensities of $(5,3,0)$ off resonance [$\AF$ in (a)] and of $(5,5,0)$ at the quadrupolar [AF $\hat{\mathcal{O}}_{22}$ in (c)] and magnetic [AF $\hat{\bm{\mathcal{M}}}$ in (b)] resonances, and the $(10,0,0)$ splitting [$\Fz$ in (a)], each scaled to the calculation at $T \to 0$.
(a) Distortions: AF $x^2-y^2$, F $z^2$ ($\Fz > 0$: elongation along the modulation axis), and F $xy$; squares: x-ray amplitudes at 6 K with their quoted uncertainties \cite{Hirai2020}.
(b) Magnetic moment $\hat{\bm{\mathcal{M}}}$ (solid) and dipole $\hat{\bm{\mathcal{J}}}$ (dashed, dimensionless), F and AF components; square: net moment inferred from the magnetization at 7 T and 2 K \cite{Hirai2019}.
(c) Quadrupoles: AF $\hat{\mathcal{O}}_{22}$, F $\hat{\mathcal{O}}_{20}$, and F $\hat{\mathcal{O}}_{xy}$.
(d) Octupoles $\hat{\mathcal{T}}^{\Gamma_4}$ (solid) and $\hat{\mathcal{T}}^{\Gamma_5}$ (dashed), F and AF components.
}
\label{Fig:tdep}
\end{figure*}

\paragraph*{Vibronic order of Ba$_2$MgReO$_6$.}
The Hamiltonian yields the experimentally observed ordered phases of Ba$_2$MgReO$_6$ (Table \ref{Table:results}, Fig. \ref{Fig:tdep}). 
In the mean-field approximation, the same phases, including the $\Fz$ elongation between $T_m$ and $T_Q$, appear for $0.42 \le f_\text{el} \le 1$, while $T_Q$ rises from 53 to 95 K and $T_m$ only from 29 to 34 K.
With $f_\text{el} = 1/2$, the ORF theory gives $T_Q = 34.7$ K and $T_m = 19.3$ K, 5\% and 7\% above the measured 33 and 18 K.
Because $f_\text{el}$ mainly sets $T_Q$, $T_m$ is the prediction of the theory.
Both deviations are comparable to the 7--9\% error of the ORF theory for the $T_c$ of classical spin models \cite{Wysin1999}.
We report the uniform (F) and staggered (AF) components on the four-site cell of the domain modulated along $z$: 
$\AF$ is the staggered $\langle \hat{\mathcal{Q}}^i_\epsilon \rangle$ and $\Fz$ the uniform $\langle \hat{\mathcal{Q}}^i_\theta \rangle$, converted to \AA\ by the factor $\ell_E = \sqrt{\hslash/(M_\text{O}\omega_E)} = 0.064$ \AA\ with the oxygen mass $M_\text{O}$, which gives the amplitudes of the normalized $E_g$ symmetry coordinates of the six O displacements of a ReO$_6$ octahedron, the $\varepsilon_v$ and $\varepsilon_u$ modes of Ref. \cite{Hirai2020}.
The calculated $\AF$ distortion agrees with experiment, the $\Fz$ elongation of 0.018 \AA\ lies within the uncertainty of the refined 0.01(1) \AA, and the zero-field net moment $m_\text{net} = 0.33$ $\mu_\text{B}$ lies below the 0.46 $\mu_\text{B}$ inferred from the magnetization at 7 T (Table \ref{Table:results}).
The calculated AF moment follows the temperature dependence of the scaled x-ray scans [Fig. \ref{Fig:tdep}(b)] \cite{Soh2024}.
The calculated $\AF$ distortion and AF $\hat{\mathcal{O}}_{22}$ follow them below $T_m$ but exceed them between $T_m$ and $T_Q$ [Fig. \ref{Fig:tdep}(a), (c)], because mean-field order parameters miss the critical fluctuations and rise too steeply below $T_Q$.

The $\Fz$ elongation above $T_m$ follows from the vibronic nature of the local states.
With the same couplings but only the ground vibronic quartet kept on each site, the mean-field $\Fz$ vanishes between $T_m$ and $T_Q$ and appears only with the magnetic order.
The admixture of excited vibronic states by the intersite fields \cite{Iwahara2023} restores it, and the quadratic vibronic coupling enhances it by a factor of 2.6 midway between $T_m$ and $T_Q$.
The $\Fz$ elongation therefore appears at $T_Q$ and grows below $T_m$, as the measured tetragonal splitting does [Fig. \ref{Fig:tdep}(a)] \cite{Hirai2020, Soh2024}; just above $T_m$ it reaches 41\% of its low-temperature value, against the measured 70\% (Table \ref{Table:results}).
Below $T_m$, the calculated $\AF$ distortion grows only slightly, as the off-resonant $(5,3,0)$ intensity does \cite{Soh2024}, whereas the $(4,1,0)$ intensity decreases \cite{Hirai2020}.
The calculated AF $\hat{\mathcal{O}}_{22}$ grows by only 2\% below $T_m$, whereas the resonant $(5,5,0)$ intensity keeps growing \cite{Soh2024} (Table \ref{Table:results}).

These remaining discrepancies, in particular the smaller $\Fz$ fraction just above $T_m$, point in part to the coupling to the uniform strain, which the lattice Green's function excludes.
The uniform $E_g$ strain couples to the octahedral distortions and to the $t_{2g}$ orbitals [Ref. \cite{Kanamori1960}, Eq. (11)] and should raise the $\Fz$ elongation between $T_m$ and $T_Q$.

Below $T_m$, the magnetic moments cant by $\phi_m = 17^\circ$ from the $[110]$ direction of the net moment, and the canted order induces a ferro $xy$ distortion of 0.001 \AA\ [Fig. \ref{Fig:tdep}(a)].
The octupoles order together with the dipoles [Fig. \ref{Fig:tdep}(d)], in line with $^{17}$O nuclear magnetic resonance (NMR), which finds sizable octupole moments \cite{Takeda2020, Takigawa2026}.

\paragraph*{Conclusion.}
A spin-vibronic Hamiltonian derived from first principles, with only the elastic coupling scaled, reproduces the quadrupole and canted magnetic phases of Ba$_2$MgReO$_6$ with the same vibronic states that describe its RIXS spectra, which supports the spin-vibronic order proposed in Ref. \cite{Iwahara2023}.
One Hamiltonian thus describes semiquantitatively the RIXS spectra and the ordered phases: both transitions, the $\AF$ distortion with the $\Fz$ elongation between $T_m$ and $T_Q$, the canted moments, and the octupoles seen by NMR.
Two extensions are the next steps toward a fully quantitative description.
Intersite correlations beyond the reaction field change the ordered phases in classical simulations \cite{Cong2026, Hart2026} and should refine the transition temperatures.
They may also explain the growth of the quadrupolar order below $T_m$ seen by resonant elastic x-ray scattering [Fig. \ref{Fig:tdep}(c)].
The coupling to the uniform strain should enhance the $\Fz$ elongation.

\paragraph*{Data availability.}
The data that support the findings of this work are provided as the Supplemental Material \cite{SM}.

\begin{acknowledgments}
The work at Chiba was partly supported by the Nippon Sheet Glass Foundation for Materials Science and Engineering.
\end{acknowledgments}

\bibliography{ref}

\end{document}